# Rapid Phase-Resolved Prediction of Nonlinear Dispersive Waves Using Machine Learning

Fazlolah Mohaghegh,[1,*] Mohammad-Reza Alam,[2] Jayathi Murthy[1]

[1]*Department of Mechanical and Aerospace Engineering, University of California-Los Angeles, Los Angeles, CA 90095, USA*
[2]*Department of Mechanical Engineering, University of California, Berkeley, California 94720, USA*

In this paper, we show that a revised convolutional recurrent neural network (CRNN) can decrease, by orders of magnitude, the time needed for the phase-resolved prediction of waves in a spatiotemporal domain of a nonlinear dispersive wave field. The problem of predicting such waves suffers from two major challenges that have so far hindered analytical or direct computational solutions in real time or faster: (i) the reconstruction problem, that is, how one can calculate from measurable wave amplitude data the state of the wave field (wave components, nonlinear couplings, etc.), and (ii) if such a reconstruction is in hand, how to integrate equations fast enough to be able to predict an upcoming rouge wave in a timely manner. Here, we demonstrate that these two challenges can be overcome at once through advanced machine learning techniques based on spatiotemporal patches of the time history of wave height data in the domain. Specifically, as a benchmark here we consider equations that govern the evolution of weakly nonlinear surface gravity waves such as those propagating on the surface of the oceans. For the case of oceanic surface waves considered here, we demonstrate that the proposed methodology, while maintaining a high accuracy, can make phase-resolved predictions more than two orders of magnitude faster than numerically integrating governing equations.

Nonlinear waves are prevalent in a number of fields such as plasma physics [1], aerodynamics of fluid-structure interaction of flying objects [2], chemical kinetics [3, 4], and solid-state physics [5]. While the prediction of nonlinear waves is already a challenging task due to the complexity of the governing equations, the problem becomes even more complicated and computationally expensive for dispersive waves i.e. when the velocity of each wave component varies with its corresponding frequency. Dispersive waves ubiquitously occur in different fields such as nonlinear optical waves [6, 7], oceanic waves [8], and flexural waves propagating in plates and bars [9]. Rapid prediction of such waves plays an import role in modeling several physical phenomena. In this work, the term rapid prediction is referred as the prediction of an event in an initial value problem before the occurrence i.e. faster than the real-time prediction. The significance of rapid predictions of waves is not limited to important physical phenomena such as weather forecasting [10], ground seismic motions in earthquake early warning [11], cargo transfer and vessel positioning in marine science [12, 13], the prediction of solar generated disturbances [14], and others [13, 15, 16]. Therefore, attaining solutions for nonlinear dispersive waves significantly faster than numerical simulations, if available, will be a significant achievement. Previous attempts have been insufficient for a faster than real-time prediction through simulations because the high computational burden of data processing and analysis precludes such predictions [17, 18].

In this letter, we focus on rapid prediction of ocean waves as an example because their accurate prediction remains as one of the most important outstanding problems in classical physics [19]. Non-linearities associated with ocean waves lead to coupled Fourier modes, necessitating complex and time-consuming simulations like broad-band high-order spectral analysis or direct numerical simulations. Furthermore, phase-resolved simulations may require a time-consuming preprocessing step. For example, in spectral analysis, in order to perform a simulation from a snapshot of the initial sea state, initial wave phases must be extracted from the data e.g. images representing the sea surface elevation. However, a wave group that appears in a still photo is a spectrum of different waves propagating in different directions which are superposed nonlinearly [20]. Therefore, a time consuming data assimilation preprocessing should be used to extract the initial conditions [21]. This extra computational burden is necessary to reconstruct the initial wave for the simulation. It is important to mention that several other applications benefit significantly from rapid predictions of nonlinear oceanic waves. For example, with an accurate rapid prediction of ocean surface height, extreme events such as rogue waves [22, 23] can be avoided or braced for. Rogue waves occur in other nonlinear wave fields such as optical waves, wind, plasma ion acoustics, Helium II superfluid second sound [24] and follow the computationally demanding general governing equations of wave evolution [52]. The other application is ocean wave energy harnessing devices that work based on the resonance of an incoming wave [25]. They have a passive spring phase control system that works based on average data of ocean surface waves i.e. sea state. If phase-resolved wave elevation can be predicated and is in hand, then with an active phase control system, wave energy convertors can tune up in real-time to work on optimum performance [26].

Along with short prediction times [27] and long computational times, typical phase-resolved ocean wave predictions work only on moderately calm or smooth standard sea states [28], where each sea state represent the sea roughness in ten different standard classes from glassy calm to phenomenal [29]. In some works, predictions are based on strong simplifying assumptions such as linear ocean waves [12, 30], or through model equations such as the weakly nonlinear Schrödinger models (NLS) [31-34]. NLS models assume a narrow-banded wave field and small steepness [33], which limits the accuracy of the prediction for broad-band dispersive waves. More accurate approaches such as fully nonlinear wave equations [35] or higher order spectral methods [36] are computationally intensive and beyond the scope of rapid predictions.

In this Letter, we employ machine learning (ML) to rapidly predict the evolution of ocean waves. Machine learning (ML) has recently emerged as a versatile data analysis tool for problems where analytical modeling is difficult because of problem complexity or where getting a solution is time intensive. Machine learning uses the results either from simulations or experiments to train a neural network algorithm, which can be used later without the necessity of solving the governing equations. ML has been used to solve a variety of complex physics problems, for example, vector vortex beam classification [37], predicting tensorial properties of atomistic systems [38], and quantum measurements [39]. ML works especially well in rapidly predicting physical phenomena: for example, a surrogate nonlinear regression model with reduced complexity can replace the solution for the molecular Schrodinger equation [40]; a neural network representation of DFT (density-functional theory) calculates forces and the energy based on atomic positions in an arbitrary size system [41]; and a neural network provides the nonlinear dependency of the logarithmic negativity of certain measurable moments [42].

Machine learning has been used to predict averaged quantities associated with ocean waves such as the spectrum and the significant wave height $H_s$ [17, 43-49], where $H_s$ is defined as four times the surface elevation standard deviation [50]. However, ML for phase-resolve analysis of ocean surface waves i.e. finding the details of surface elevation as a function of space and time, is still not in hand: limitations include calm to moderate sea states [51-53], too short prediction periods [16], requirement for a time-consuming preprocessing [16], and unrealistic simplifications such as the assumption of a fully developed sea [51]. To address these issues, this work proposes a novel approach toward the short-term rapid forecasting of ocean waves. Using machine learning techniques, we develop a method which is at least two orders of magnitude faster than even efficient spectral analysis [36] direct simulations, leading to a faster than real-time calculation of incoming waves.

The implementation of a machine learning technique includes preparation of the training set, setting up the algorithm, training the network parameters, saving the trained network for the rapid prediction, and testing its accuracy. To prepare training and testing sets, we use direct simulations to provide data of wave elevation evolution with time. The simulation assumes an incompressible, inviscid, and irrotational flow based on the formulation described by Liu and Yue [54], which employs a nonlinear high-order spectral method based on perturbation theory [36]. The simulations are based on a range of random initial phases over JONSWAP spectra, which describe the energy distribution with frequency over the ocean surface [55]. Therefore, the wave elevation from each simulation starts with the amplitudes of the wave components obtained from JONSWAP spectra but with different random initial conditions, leading to different wave elevation histories.

The training set is constructed with the results based on a polychromatic incident wave in one direction with the length $4.5 \ km$. Using a $4^{th}$-order nonlinearity, we simulated the wave evolution in an ocean with a depth of $300 \ m$ and in the sea state 6 i.e. very rough [29]. Due to the differences in the wave patterns at different sea states, we employ a different network for each sea state, which is already known as an average value. Discretizing the spatial domain using 1024 discrete nodes, and using 32 time steps per JONSWAP standard peak period [55], a spectrum consisting of 100 wave components constructs the wave field based on random initial phases for each component. We collect and save the wave height time-history data for 200 peak periods [55], where the peak period stands for the wave period corresponding to waves carrying the most energy in the whole spectrum. The database contains 20 runs related to different initial conditions. Each training epoch randomly picks one of these runs and randomly selects a place in the time series data for that run. At the selected place, the network uses the wave height data from last 1000 time steps and predicts the wave height for the next 1000 time steps, corresponding to an elapsed time of $303.2 \ s$. The data at each time step contain the wave elevation at 1024 physical points and may or may not include a rogue wave. Rogue waves are defined as big waves with crest-to-trough heights more than $2H_s$ and crest heights more than $1.25H_s$ [56]. A total of 50,000 training epochs performs the training, with batches of size 5. The mean squared error between the simulation data and the prediction is used to calculate the loss function.

We employ a supervised generative machine learning technique i.e. a recurrent neural net (RNN) based on surface wave elevation data. Recurrent neural networks employ an internal state as a memory in a time-series data. This state variable enables the network to exhibit temporally evolving behaviors and to construct future data based on its history. Here we use sequence to sequence (seq2seq) [35] RNN technique that operates as an autoregressive model with minimum assumptions on data with sequence structure. Autoregressive models employ previous data which are of the same type as the predicting data. Therefore, in this analysis we only use surface elevation values as the input data and disregard the explicit impact of other parameters on the wave height such as wind speed, seabed topology and so on. Thus, the effect of these parameters on wave evolution is implicit in the history of surface elevation. Hence, the ML network takes a timeframe i.e. a time window of input nodes representing the history of wave elevation for one simulation and then outputs a timeframe of the same size predicting future wave elevation.

In contrast to standard RNNs in which there is a single value for every input node, each node in the implemented network contains a vector of surface wave elevation data in the domain (Fig. 1). All vectors have the same number of features, i.e. the spatial resolution is fixed in the entire data. Here, the traditional convolution operation [57] is not used to extract features. In traditional convolutions, the connection between the data i.e. sequential time instances for this case, is enforced through convolutional kernels and filters. Consequently, mapping a data patch with a traditional convolution operation leads to arbitrary complexity [58]. The complexity determines how accurately the model performs on the training data. Thus, if the model becomes too simple, the network cannot be trained for accurate predictions. To avoid this problem and to include vectorized input for each node to the network, instead of the traditional convolution, a new type of convolution designed for the time-series data is used. We implement a modified version of the convolutional recurrent neural net (CRNN) [58] which directly uses consecutive frames of wave elevation in the domain as a patch of data to extract spatiotemporal features. The CRNN approach assumes that the data at each time step are a continuous and smooth transition from the previous time step, and that transitions in space are also continuous and smooth. Therefore, the prediction of one spot in the space depends also on the other spots and time steps. As a result, CRNNs can reconstruct spatiotemporal variations with great accuracy.

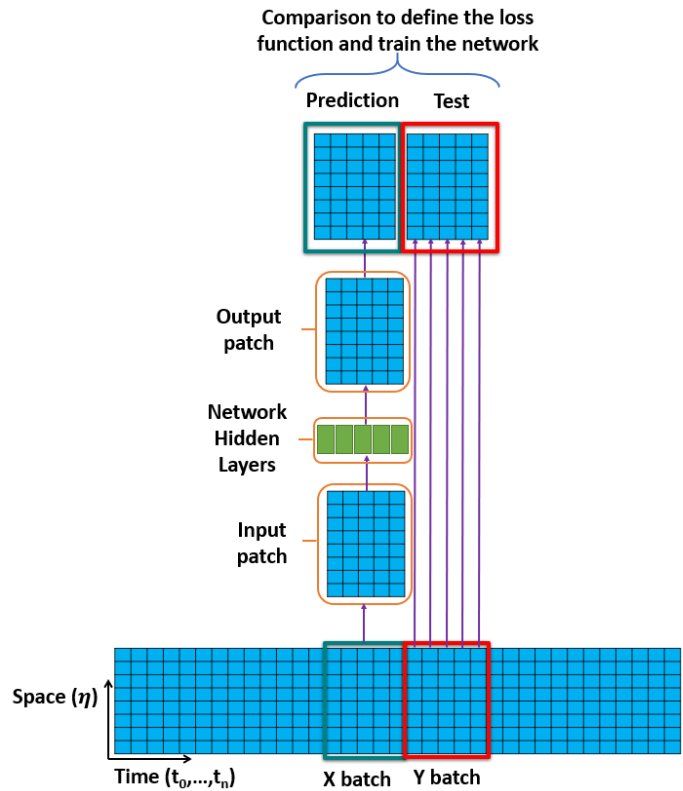

FIG 1. CRNN network for a spatiotemporal domain: Each node contains wave elevation data in the whole physical domain. The input to the network is wave elevation data for the last 1000 time steps and the output is the prediction of wave elevation for the next 1000 time steps.

While the original CRNN [58] applies pooling by using strides over the patch, the current work does not implement strides to preserve features related to sudden and rapidly disappearing spikes in the ocean surface. Instead, we consider the whole spatiotemporal patch of input data as one encapsulated package and predict the data associated with the next spatiotemporal patch of the same size as the output. Furthermore, in the forward propagation phase, the network takes a random patch from the spatiotemporal domain i.e. a timeframe (window) from the time-series data and generates a sequence of hidden states to encapsulate features. Then, by using the extracted features, the network reconstructs the patch of the spatiotemporal domain. The training process is similar to the training used in a standard RNN method, i.e. backpropagation through time (BPTT) [59]. The Adam optimization [60] technique trains the network over known time series data by comparing ML results with known results at all instances in the prediction timeframe. The output is a timeframe of nodes representing a set of wave elevation values in the physical domain. After the training, the network is saved to be used to predict the wave elevation field.

In order to test the performance of the network, we compare our results with test cases obtained with initial conditions different from the training cases. We consider the problem of rogue wave prediction and select one of the simulations that contains a rogue wave. Then we select up to 2000 time steps before the rogue wave occurrence and use the first 1000 time steps as the input to the network. The prediction time frame $T_{tf}$, i.e. the next 1000 time steps, includes a rogue wave and these data are used as the test case. Recall, however, that the network is trained with simulation data that may or may not contain rogue waves. The trained machine learning network with forward propagation predicts the future wave evolution in the domain. The machine learning prediction takes 9.8 $sec$ on an Intl(R) Core(TM) i5-7200 computer with a 2.7 GHz CPU and 8 GB memory. This time includes the entire processes related to the prediction, i.e., reading and loading the saved network data, network forward propagation and postprocessing. In contrast, the direct simulation of 1000 timesteps takes 528.3 $sec$ if we continue the simulation after the first 1000 timesteps. Comparing with the ML prediction time of 9.8 $sec$, the predictability horizon is 518.5 $sec$ and the real-time ratio i.e. the ratio of the prediction time to the time of occurrence is 0.0186. Needless to mention that using simulations requires the expensive preprocessing task of data assimilation and extracting the initial phases and amplitudes from the spectrum of wave elevation data. As this process can be even longer than the simulation time [61], the modeling can take significantly longer than 528.3 $sec$.

The results shown in Fig. 2, compare the performance of the machine learning network at different instances in the prediction timeframe $T_{tf}$, which has 1000 timesteps. Comparisons are for 0,200, 400, 600, 800 and 1000 steps, and correspond to 0, 0.2 $T_{tf}$, 0.4 $T_{tf}$, 0.6 $T_{tf}$, 0.8 $T_{tf}$ and $T_{tf}$ show how the wave evolves to a rogue wave, which occurs at a time step of 1000. Clearly, the machine learning algorithm can accurately predict the evolution of the wave profile. To quantify the error, we use the normalized $L1$ norm as follows:

$$L1 = \frac{\sum |\eta_{simulation} - \eta_{ML}|}{n_x \, H_s} \tag{1}$$

In the Eq. 1, the summation is applied to every spatial grid point at each time instance. In this definition, the sum of the difference magnitude between the simulation and ML is normalized with the significant wave height $H_s$ times the spatial resolution $n_x$. The error analysis leads to an average error of 5.5% at the time step at which the rogue wave occurs. The algorithm can also predict the time and the place of a rogue wave with great accuracy; the error in the spatial location of the wave peak is one spatial grid spacing and the temporal error is one time step. However, the method predicts the wave height at the rogue wave peak with about 27.8% error, which is lower than the overall accuracy. This is because a rogue wave is an extreme and rare event. Therefore, when a network is trained based on the general situation, there will be some discrepancy for rare events. Despite this error, this rogue wave prediction is still very helpful since it signals rogue wave formation, but it may lead to ignoring the danger when the wave height is close to the threshold of the rogue wave height criterion.

For a better comparison, we use Fourier analysis to decompose the wave into discrete frequencies and compute the power spectral density (PSD). To perform this analysis, we first extract the time history of wave elevation at each physical point in the domain, as shown in Fig. 3 for a point in the middle of the domain and for the location where a rogue wave occurs. In Fig. 3, the peak of the rogue wave occurs at 0.8 $T_{tf}$. A Fast Fourier Transform (FFT) is applied to all spatial points of the wave elevation data for ML and simulation separately. Then resulting FFT is normalized with the frequency. We average the value of PSD over all 1024 spatial nodes in the domain. The results, shown in Fig. 4, indicate that the ML prediction compares well with the direct simulation. The deviation of both ML and direct simulation data from the JONSWAP spectrum used in our initial conditions is due to nonlinear interactions that occur in the wave spectrum during wave evolution.

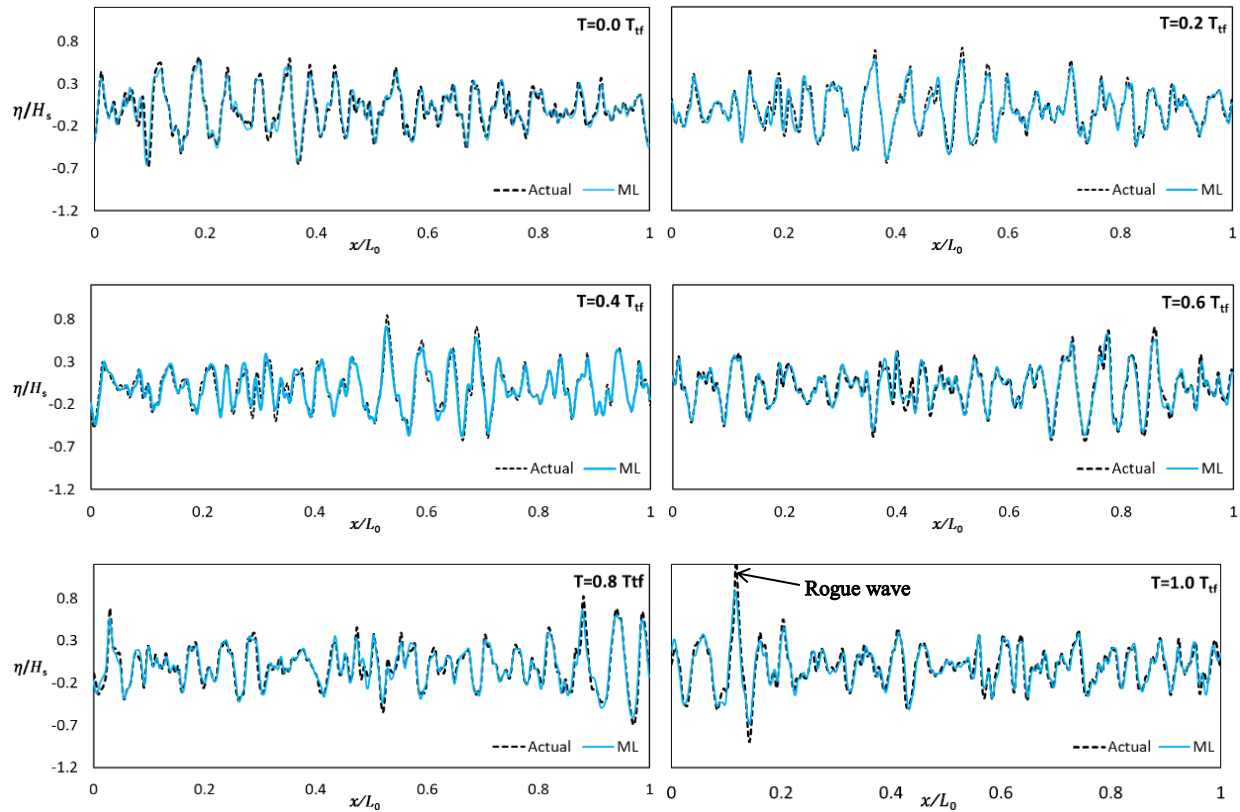

*FIG 2. Comparison between the direct simulation and machine learning in the prediction of ocean waves for sea state 6 i.e. very rough sea state. The rogue wave occurs in the last figure at time step 1000 i.e. $T = T_{tf}$. The values on the horizontal and vertical axes are normalized with the domain size of $L_0 = 4.5\ km$ and with the significant wave height $H_s = 5.0\ m$, respectively.*

In summary, we have proposed a revised version of a convolutional neural network (CRNN) to accurately predict nonlinear dispersive wave evolution including rogue waves. For the case of oceanic rogue waves, the timeframe for prediction in the present simulations allows an advance warning of about five minutes, which is sufficient for ships to take protective action. We demonstrate that our method predicts ocean surface wave elevations accurately in a domain of size $\mathcal{O}(5\ km)$. Moreover, the network forecasts the entire wave field and can therefore predict the location and timing of rogue waves. Furthermore, our technique may also be used on actual sequences of ocean surface images obtained from radar or other data acquisition systems to make predictions of future evolution.

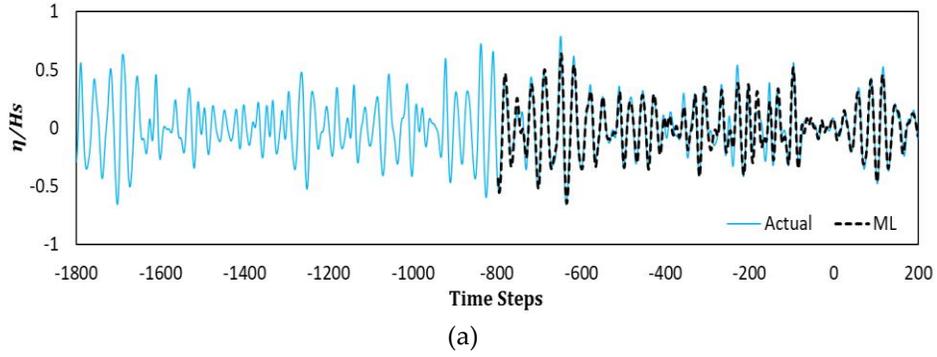

(a)

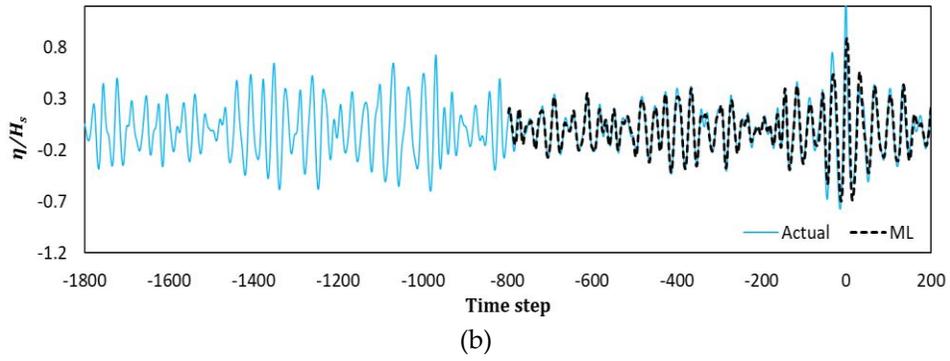

(b)

*FIG. 3- History of wave elevation at the middle of the domain for one of the cases in sea state 6 i.e. very rough sea state. The results from machine learning are compared with direct simulation for 1000 timesteps for (a) a point in the middle of the domain (b) the place that the rogue wave occurs, $0.1172L_0$ from the left side of the domain. The rogue wave occurs at time step zero and $0.8\ T_{tf}$. The first 1000 time steps in each figure is the input to the network and the next 1000 time steps are the generated output.*

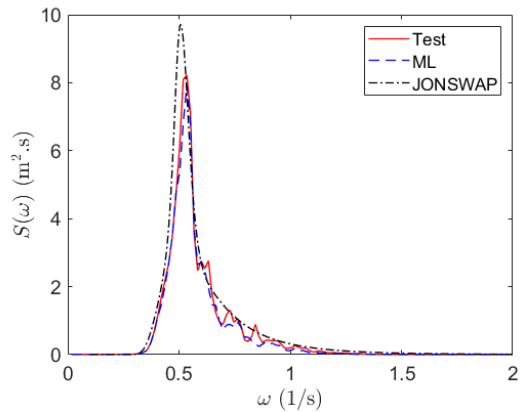

*FIG 4- Comparison of average power spectral density of the ML prediction and that from direct simulation at sea state 6 i.e. very rough sea state as a function of frequency, $\omega$. Also shown is the power spectral density corresponding to the JONSWAP spectrum used to initialize the direct simulation.*

*ehsanm@ucla.edu